# Rapid Conformational Analysis of Semi-Flexible Liquid Crystals

J. L. Hobbs, [1] C. J. Gibb, [2] E. Cruickshank, [3,4] R. Walker, [3] R. J. Mandle [1,2,*],

[1]School of Physics and Astronomy, University of Leeds, Leeds, LS2 9JT, UK
[2]School of Chemistry, University of Leeds, Leeds, LS2 9JT, UK
[3]Department of Chemistry, School of Natural and Computing Sciences, University of Aberdeen, AB24 3UE, UK
[4]School of Pharmacy and Life Sciences, Robert Gordon University, Aberdeen, AB10 7GJ, UK
[*]r.mandle@leeds.ac.uk

**Abstract**

We present an approach for rapid conformational analysis of semi-flexible liquid crystals. We use a simple graphical user interface (GUI) tool that leverages rules-based methods for efficient generation of bend-angle distributions, offering a significant improvement over traditional single-conformer analysis. Our methods demonstrated proficiency in approximating molecular shapes comparable to those obtained from molecular dynamics (MD) simulations, albeit with notable deviations in the under sampling of hairpin conformations and oversampling of extended configurations. Re-evaluation of existing data revealed an apparent weak correlation between $N_{TB}$ transition temperatures and bend angles, underscoring the complexity of molecular shapes beyond mere geometry. Furthermore, we integrated this conformational analysis into a pipeline of algorithmic molecular design, utilizing a fragment-based genetic algorithm to generate novel cyanobiphenyl-containing materials. This integration opens new avenues for the exploration of liquid crystalline materials, particularly in systems where systematic conformer searches are impractical, such as large oligomeric systems. Our findings highlight the potential and growing importance of computational approaches in accelerating the design and synthesis of next-generation liquid crystalline materials.

**Introduction**

Virtually all low-molecular weight liquid crystalline materials are flexible molecules that can adopt multiple conformations. Meyer and Dozov independently theorised that molecules with a bent-shape could spontaneously form a nematic-like phase that has a helicoidal structure and thus is locally chiral, even when formed of achiral molecules. [1] This is widely known as the twist-bend nematic ($N_{TB}$) phase, and its experimental observation in 2011 is a landmark work. [2] The importance of molecular shape within the context of the $N_{TB}$ phase has been appreciated for some time; consider that dimers of odd-parity have a strong predisposition to exhibiting this phase that is not shared by their even-parity homologues. [3]

The $N_{TB}$ phase is the "structural link" between the classical nematic phase and the (helical) cholesteric mesophase [4]. A number of techniques have measured, [4-9] inferred, [10,11] or simulated [12] the pitch length of the $N_{TB}$ phase to be on the order of a few nanometres (although material dependent). It also seems reasonable to suppose that more recently discovered twist-bend smectic phases will also exhibit a dependency upon molecular shape. [13-19] Moreover, the potential existence of polar variants must also be seriously considered, [20] given the transformative discovery of conventional polar-ordered nematics. [21-25] Very recently, polar heliconical phases reminiscent of the $N_{TB}$ have been reported. [26,27]

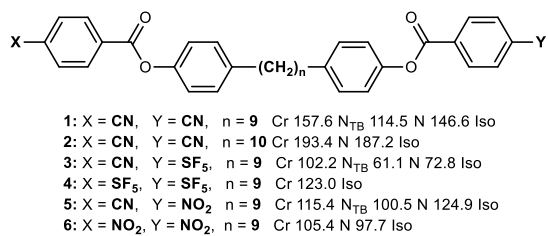

1: X = **CN**, Y = **CN**, n = **9**   Cr 157.6 N$_{TB}$ 114.5 N 146.6 Iso
2: X = **CN**, Y = **CN**, n = **10**  Cr 193.4 N 187.2 Iso
3: X = **CN**, Y = **SF$_5$**, n = **9**   Cr 102.2 N$_{TB}$ 61.1 N 72.8 Iso
4: X = **SF$_5$**, Y = **SF$_5$**, n = **9**   Cr 123.0 Iso
5: X = **CN**, Y = **NO$_2$**, n = **9**   Cr 115.4 N$_{TB}$ 100.5 N 124.9 Iso
6: X = **NO$_2$**, Y = **NO$_2$**, n = **9**   Cr 105.4 N 97.7 Iso

**Figure 1:**   Transition temperatures (T, ° C) of some liquid crystalline dimers with varying spacers and terminal groups [28-30]

We caution that molecular shape does not dictate the formation of the N$_{TB}$ in isolation from all other factors; the propensity of various functional groups to promote or destabilise liquid crystalline order must also be considered, as exemplified by considering the materials shown in Figure **1.** The change from odd (compound **1**) to even (compound **2**) parity spacer has the expected effect, leading to the loss of the N$_{TB}$ phase. However, as does replacement of the nitrile units with either nitro (NO$_2$; **6**) or pentafluorosulphanyl (SF$_5$; **4**). A mix of one nitrile and one NO$_2$/SF$_5$ group retains the N$_{TB}$ phase (**3**, **5**), with diminished onset temperature relative to the parent material. [28,29]

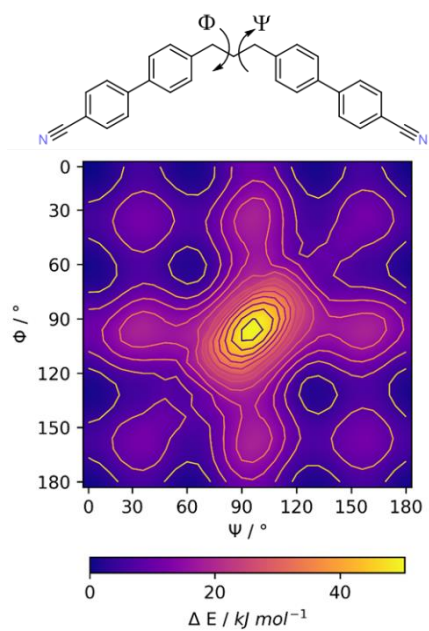

**Figure 2:**   Potential energy surface of CB3CB generated by relaxed scans about the two indicated dihedral torsions (Psi, Phi) using the PM3 semi empirical method

Molecular shape is intimately linked with the formation of liquid crystalline phases; [31] an accurate description of molecular "shape" for such materials relies on considering an ensemble of conformers which are energetically accessible at some a given temperature. For small molecules with limited conformational freedom (e.g. CB3CB in **Figure 2**) it is possible to scan over the potential energy surface using semi-empirical or even DFT methods, although for larger and/or more flexible systems systematic conformer searching can become impractical. Molecular dynamics (MD) simulations are capable of generating liquid crystalline order; [32] These directly provide conformational ensembles for a given force-field and could be considered as the 'gold standard' for computational assessment of molecular shape, as the influence of the local environment is also captured. While MD simulations offer detailed insight into conformational distribution in liquid crystalline materials, it is most common for a single minimum (or low) energy conformer to be generated using electronic structure calculations (e.g. DFT, MP2) which is assumed to representatively encompass all energetically feasible conformers of the molecule. This assumption streamlines the computational process, and although this may not capture the dynamical complexity exhibited in MD simulations (or the reality they simulate), such practices reflect a compromise in terms of time cost, and knowledge gained. Fast conformer generation algorithms have made impressive progress in their ability to replicate experimental conformations, [33] especially when using rules and/or knowledge based methods. [34-45] Stochastic sampling of conformational space allows study of highly flexible or large molecules [46-51] which suffer a combinatorial explosion when treated with systematic methods. [52]

Application of these fast conformer-generation methods is thus of interest as a means to assess the shape of liquid-crystalline materials, and herein we describe assessment of the use of rules-based methods for this purpose. Rule (or knowledge) based methods use a library of experimentally derived data for torsional angles, ring conformations; input molecules are broken into smaller fragments, and conformations of these extracted from the data library either stochastically or systematically. [48] Other fast-conformer search methods applied to liquid crystalline materials include Monte Carlo searches, [53] stochastic MD, continuous torsional potentials, [2] systematic rotor searches, and genetic-algorithm searches. [54]

Firstly, we evaluate the performance of rules-based methods in reproducing the MD-derived bend-angle distribution of the well-known $N_{TB}$ material CB7CB, and select a baseline methodology for subsequent studies. Secondly, we use rules-based methods to re-examine the bend-angle distributions of families of liquid crystalline dimers where conformer data has been reported previously. Thirdly, we use these methods to examine materials of special interest where conformer data is lacking – dimers, $N_{TB}$ tetramers, and even-parity $N_{TB}$ materials. Finally, we apply our learnings to a set of materials constructed via a fragment-based genetic algorithm for generative molecular design. This allows us to identify materials that are structurally distinct from, yet comparable in terms of gross-shape to conventional $N_{TB}$ systems, suggesting possible future synthetic targets.

**Methods**

*Conformer GUI Tool*

The Rdkit [55] software package contains a number of in-build rules-based methods that can be utilised for the generation of conformers. These can take advantage of knowledge of basic chemistry, of torsional preferences (etc.) embedded within the various rules-based methods.

We considered that the ability to perform rapid conformational analysis of semi-flexible liquid crystals could be especially attractive to those working to synthesise such materials, and therefore we have built a lightweight GUI based program that enables the user to generate conformer libraries and bend-angle distributions (available from https://github.com/RichardMandle/conformer). The user can load a .mol file (which can be generated by virtually any chemical drawing software package), or supply the molecule as either SMILES or SMARTS if preferred. If bend-angle distributions are required to be calculated then the user can either select pairs of atoms that define the vectors of interest, or can pass SMILES or SMARTS strings to look-up relevant functional groups (e.g.

cyano/nitrile, C#N). The user selects the conformer generation engine, the number of conformers to generate (other options allow finer control over this) and the conformer library is then generated. However, we realise that simply being able to generate an arbitrary number of conformers might be useful beyond liquid-crystalline dimers, and so the user can select 'None' to bypass the angle calculation steps. After generating a set of conformers the bend-angles can be recalculated for a different set of reference vectors.

Additionally, this program provide s a link to the widely used Gaussian suite of programs to enable electronic structure calculations to be performed; for example, evaluation of the energy of each conformer using available methods. Once complete, the user can access and export information and coordinates of the generated conformer library. Conformer probabilities are determined from the energy relative to the minimum energy conformer using Boltzmann statistics, assuming a temperature of 298 K (unless otherwise stated). The bend-angle distribution is displayed, and other 3D descriptors (e.g. radius of gyration) are also available; all data can be exported as .csv. The user can view and export any and all conformers as .sdf which permits visualisation in external software packages, and save/load the current interactive conformer generation session.

*MD simulations and Electronic Structure Calculations*

We performed MD simulations of GAFF force field in Gromacs 2019.2 with GPU acceleration through CUDA. [56-62] Force field parameters were generated using RESP [63] charges from B3LYP/6-31G(d) [64,65] calculations using the Antechamber tool in Amber20 [66,67] and converted to Gromacs .itp format using Acpype. [68] Energy minimisation using steepest decent and sequential equilibration in the NVE, NVT and NPT ensembles afforded an isotropic configuration which was used as a starting point for further production simulations. Production MD simulations were performed using an anisotropic barostat (Parinello-Raman, 1 Bar), [69,70] a Nose-Hoover thermostat [71,72] and were performed for in excess of 1 microsecond using a timestep of 0.2 fs. The second rank orientational order parameter (<*P*2>) was calculated *via* the Q-tensor using MDTraj. [73] All electronic structure calculations were performed using Gaussian G09 revision d01. [74]

**Results**

For the first point of comparison we sought to demonstrate that the rules-based methods can yield conformer libraries which are comparable to those derived through electronic-structure calculations and MD simulations, these being the current state-of-the-art. We selected CB7CB as a system for our initial test as it is arguably the most widely studied $N_{TB}$ system.

The MD simulation of CB7CB was performed at a temperature of 363K and for 800 molecules in the $N_{TB}$ phase for over one microsecond. There was not a significant difference in conformation when the simulation was in a nematic configuration at higher temperature (430 K), and our results in both phases are in keeping with those reported earlier by Wilson et al. [12] Data for the conformational landscape using the rotational isomeric state approximation with energy evaluated using the AM1 semi-empirical method (RIS-AM1) was taken from ref [75].

For comparison we used the ETKDGv3 (Experimental-Torsion with Knowledge Distance Geometry version 3) method to generate 2048 conformers (the selection of the rules-based method and number of conformers will be discussed shortly). We define the bend-angle as the angle between the vectors defined by the C and N atoms of each nitrile group within CB7CB, while the probability is determined from the relative energies of all conformers. From the list of angle/probability pairs we compute the histograms shown in Figure **3**. The ETKDGv3 method gives results that approximate the conformational distribution from MD with some caveats. Firstly, the population of 'hairpin' conformers with small bend-angles (arbitrarily chosen here as <60 degrees) is slightly undersampled (10.5% for MD versus 8.1% for ETKDGv3). Secondly, the population of fully extended configurations is overestimated when compared to MD, giving a higher population and a narrower distribution of bend-angles. Computing energy of the ETKDGv3 geometries with AM1 is slower (as each conformer has a single point energy calculation performed) and only marginal improvement.

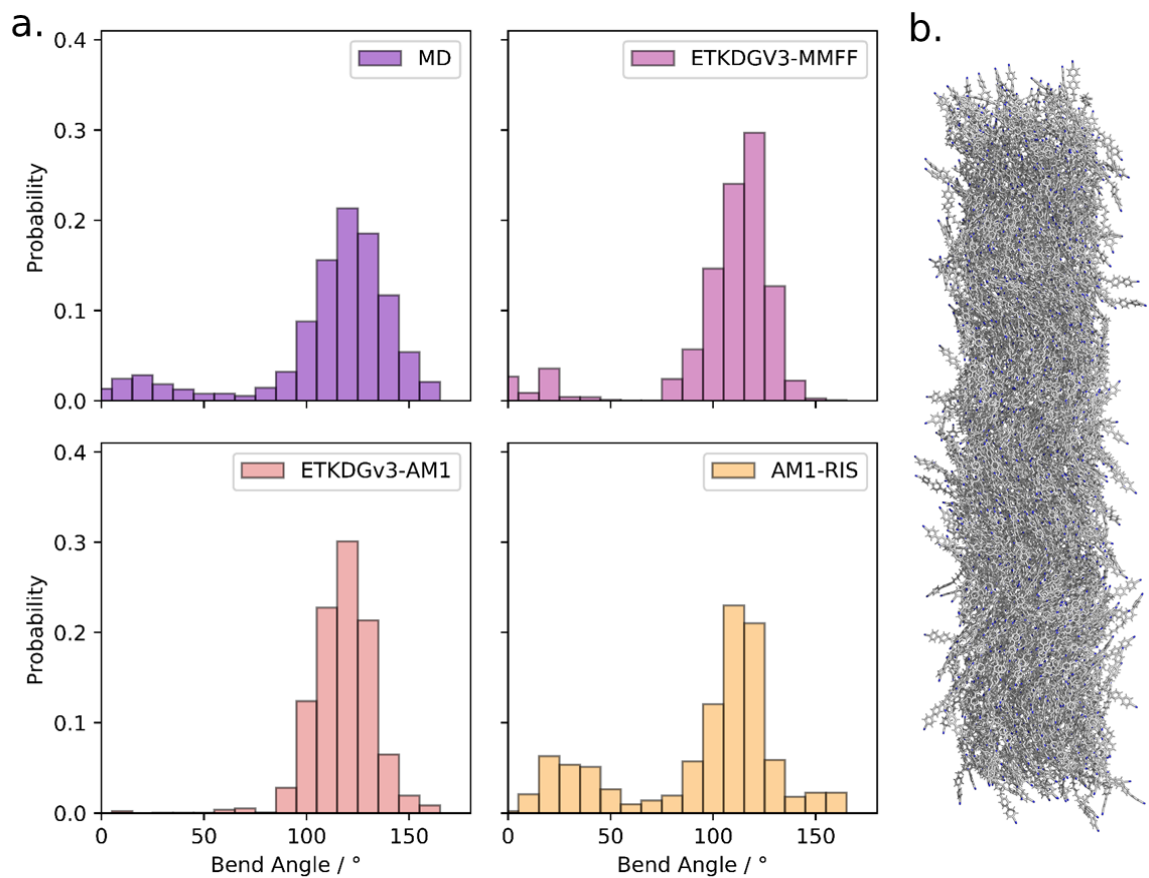

**Figure 3:** (a) Comparison of the performance of bend-angle distributions of CB7CB that result from calculation *via* different methods: MD, ETKDGv3 with MMFF and AM1 energy evaluation, rotational isomeric state (RIS) approximation with AM1 optimisation. (b) Instantaneous configuration of an MD simulation of CB7CB at a temperature of 363 K performed as part of this work.

The bend-angle distribution that results from any rules-based method neglects the influence of the local-environment in the condensed phase and so will inevitably differ to that obtained in atomistic molecular dynamics MD, where it is considered explicitly. However, the advantage here is that ETKDGv3 method is remarkably fast, with the aforementioned results generated in a few minutes using a single CPU core (Intel i7-1187G7) on a modern laptop computer. Contrast this with the MD simulations for which the production runs alone each took circa 7 days on a modern GPU (Nvidia V100, 5120 CUDA cores). A second advantage results from this speed; whereas screening hundreds of different molecules is essentially trivial with rules-based method and, as will be demonstrated later, this is sufficiently rapid to enable interfacing with algorithmic tools for generative molecular design.

We performed various screens to identify optimal parameters for conformer generation; the effects of various options are presented in the SI to this article (Figure SI-1, SI-2, SI-3, SI-4). Briefly, we found that ETKDGv3 method to generate an initial library of 2048 conformers that we prune with a RMSD threshold cut-off of 0.675 gave good results (and, typically, a library size in excess of 500 conformers after pruning). These were therefore set as the default options in our GUI tool and, unless otherwise stated, are used from hereon. Having established an optimal set of parameters for conformer searching, we now revisit two families of CBnCB dimers with varying linking groups for which information on bend-angle distributions has been reported, [75,76].

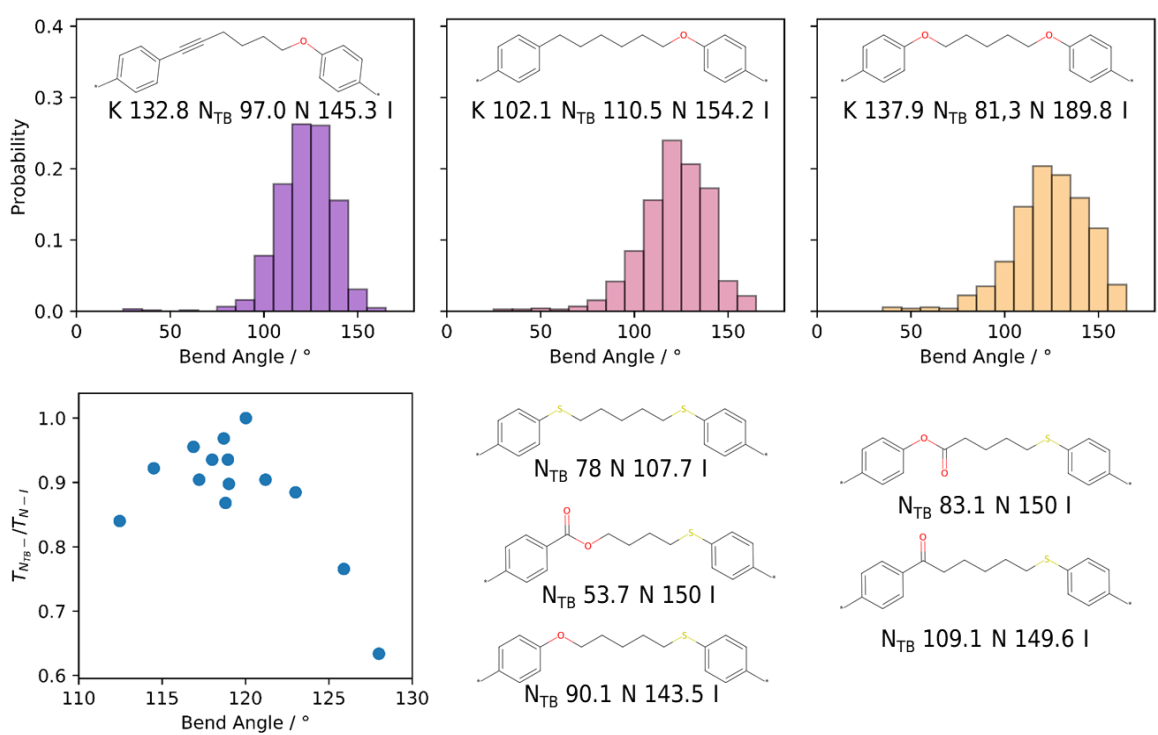

**Figure 4:** Histogram plots of bend-angle probabilities for the cyanobiphenyl dimers with heptamethylene-equivalent spacers reported by Archbold *et al.* [75] Transition temperatures (°C) were taken from Archbold *et al.* Data for additional materials shown in the figure were taken from refs [76,77]. For clarity, the 4-cyanobenzene portion of the mesogenic units is denoted as "[*]".

The average bend angles obtained using the ETKDGv3 method are larger than those in the original papers and the distributions obtained are notably broader. The data follows the expected trend, with the alkyne/ether linked system having the narrowest average bend angle, and the ether/ether linked material having the largest. We computed the bend-angle distribution for all materials reported in ref [75] as well as the additional thioether linked materials shown in **Figure 4** (transition temperatures taken from ref [77]). The scaled transition temperature ($T_{NTB-N} / T_{N-Iso}$) does appear to have a relationship with bend-angle, although this is also entangled with the propensity for a given molecular structure to exhibit the $N_{TB}$ phase. In other words, molecular shape is clearly important, but cannot be taken in isolation from other factors, as shown earlier in **Figure 1**. We would also add that the scaled transition temperatures themselves reflect many biases, for example: bounds imposed by working temperature ranges; limits of chemical stability; small sample sizes; similar material types. The relationship between scaled transition temperature and bend-angle should be treated cautiously for this reason, and would greatly benefit from additional data and new materials.

The $N_{TB}$ phase is not only observed in liquid crystalline dimers, being found in a few oligomeric materials that are expected to have more complex conformational landscapes. The alternating pattern of rigid mesogenic units and flexible spacers leads to cumulative effects; rather than considering the mesogenic units as rotationally decoupled it is more accurate to say that the influence of one mesogenic unit upon any other is directly related to their topological proximity. As the separation between two units grows, their influence on one another diminishes with the result that the preference for a given bend angle diminishes also. The tetramer $T4_9$ [78] was one of the first liquid crystalline oligomers reported to exhibit the $N_{TB}$ phase and features four mesogenic units in a linear sequence separated by nonamethylene spacers. A systematic conformational search of $T4_9$ would be prohibitively expensive, and stochastic methods are necessary. Owing to the large number of rotatable bonds in $T4_9$ we used a larger number of conformations; after removing identical entries we were left with 66k unique conformers.

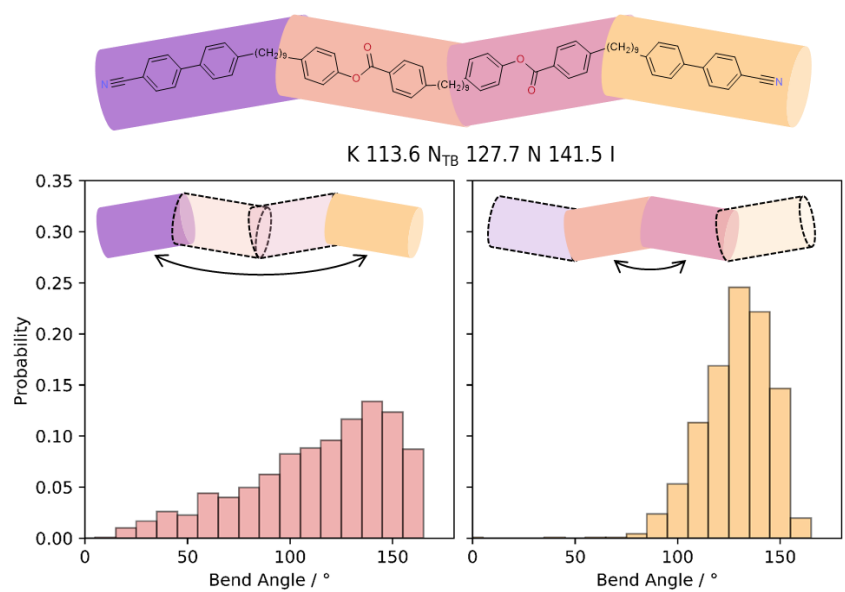

**Figure 5:** Molecular structure and transition temperatures (° C) of the tetramer $T4_9$; [78] along with bend-angle distributions between the four mesogenic units.

As shown in **Figure 5** we find the bend angle between adjacent units to be comparable to that of low molecular weight materials; with an average bend-angle of 125 ° and a comparable distribution of angles to dimeric materials. Compared to the CBnCB materials the distribution is broader in $T4_9$, owing to the ability of the carboxylate ester to adopt different conformations. We find the distribution of angles between the outermost mesogenic units is significantly broader than for adjacent core units for the reasons discussed above.

Now consider tetramers of more radical structure, which feature variations in the parity (and therefore also the length of) the spacer units. [79] We focus on two materials, one with the parity sequence LBL (linear-bent-linear) and one with BLB (**Figure 6**). Both materials exhibit the $N_{TB}$ phase with although the compound with two odd-parity spacers exhibiting a higher scaled $N_{TB}$ transition temperature (defined here as $T_{NTB-N}$ / $T_N$-Iso) than the material with just one odd-parity unit. The odd-parity spacer(s) contribute a bent molecular shape that is (mostly) preserved by the even-parity spacer(s), which is the origin of this difference in mesomorphic behaviour.

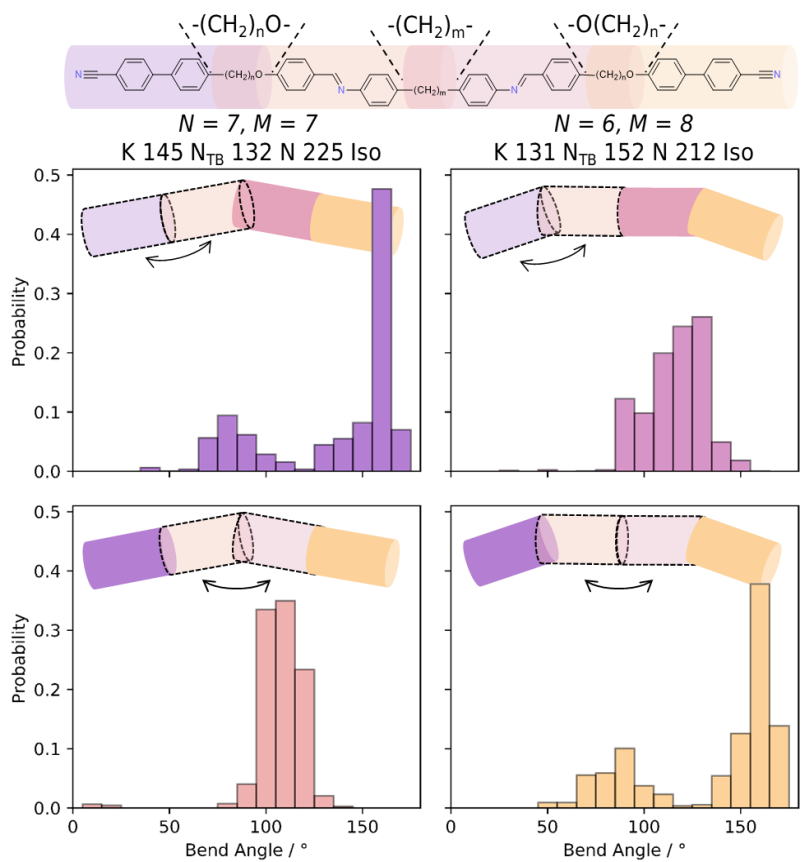

**Figure 6:** Molecular structure and transition temperatures (° C) of two tetrameric materials with linear-bent-linear (left) and bent-linear-bent (right) topologies, [79] along with bend-angle distributions between pairs of adjacent mesogenic units. The scaled NTB/N transition temperatures are 0.80 (left) and 0.81 (right).

The specific molecular structure that generates the bent-shape need not be an odd-parity spacer unit; consider the piperazine containing bent-core material UD68 (Figure SI-6). [80] Imrie et al recently reported the first liquid crystalline dimer that has an even parity spacer (i.e. the longest linear sequence of non-hydrogen atoms is an even number) yet exhibits the $N_{TB}$ phase. [81] From this result we can infer that the material has a bent shape, and from basic chemistry we expect that simple disulphides will have a torsional angle of circa 90 °. We generated a library of conformers using the ETKDGv3 method for this disulphide-bridge dimer along with some relevant example members of the CBnCB family. When the central spacer is short, there is a marked difference between even- and odd- parity dimers, as exemplified by CB6CB and CB7CB (also, CB4CB through to CB11CB in the ESI to this article, Figure SI-5). Compared to CB17CB, the disulphide bridge dimer has a slightly smaller average bend angle and a slightly narrower distribution of angles. The inclusion of a disulphide bridge has much the same effect as a single *gauche* conformer; one might expect therefore that for sufficiently large central spacers the even parity CBnCB materials will also exhibit the $N_{TB}$ phase (as the probability of a single *gauche* in the chain becomes dominant).

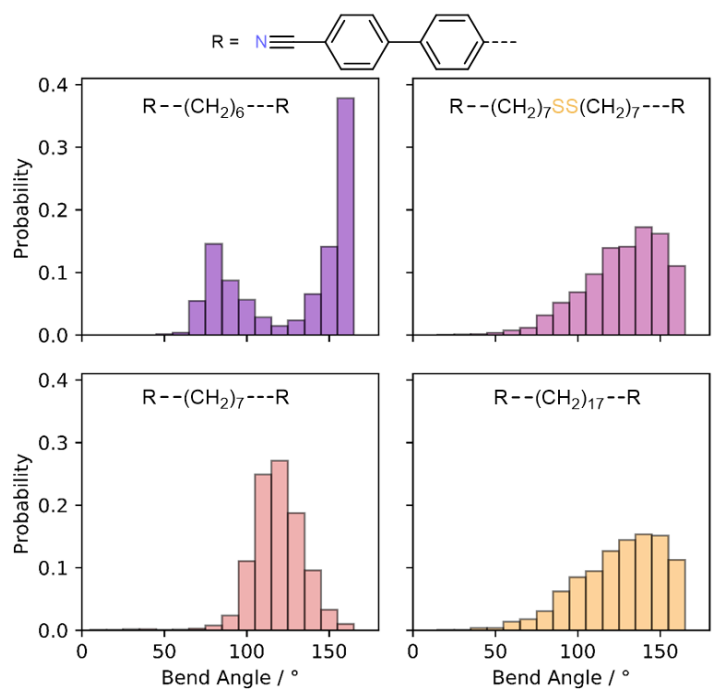

**Figure 7:** Bend-angle distributions for CB6CB, CB7CB, CB17CB, and CB7SS7CB computed with the ETKDGv3 method as described in the text.

The material RM1081 (also called FBO11ODFCB3) has the unusual phase sequence N-$N_{TB}$-SmA-SmX; [82] the $N_{TB}$ phase existing for a short temperature interval before yielding a smectic A phase, the implication is that the $N_{TB}$ helix forms then readily unwinds. Curiously, even minor changes to molecular structure eliminate the phase sequence (e.g. variations in central spacer length, terminal chain length, mesogenic unit type, linking unit types, fluorination pattern). [83] We generated a 2048 conformer library of RM1081 using the rules-based ETKDGv3 method. This affords a relatively broad distribution of bend-angles centred on ~ 130 °C, although this is in keeping with other ether-linked dimers studied in this work (e.g. CBO7OCB). Separately, fully atomistic MD simulations of RM1081 were performed using the GAFF force field, variously yielding nematic and $N_{TB}$ phases - we do not observe lamellar phases forming over the duration of the simulation, which is circa 1 microsecond. As we are principally interested in the $N_{TB}$ phase we focus only on this phase, which we find to have the bend-angle distribution given in **Figure 8** with an average bend angle of ~ 125 °. To a first approximation the rules-based method reproduces the bend-angle distribution obtained from MD with two important caveats. Firstly, the population of hairpin conformers is significantly underestimated in this system – MD predicts ~ 6% of molecules of RM1081 have a bend angle of less than 60 °, whereas the ETKDGv3 method predicts <0.5%. Secondly, the bend-angle distribution is narrower for MD, because of the neglect of the nematic environment being in the rules-based method.

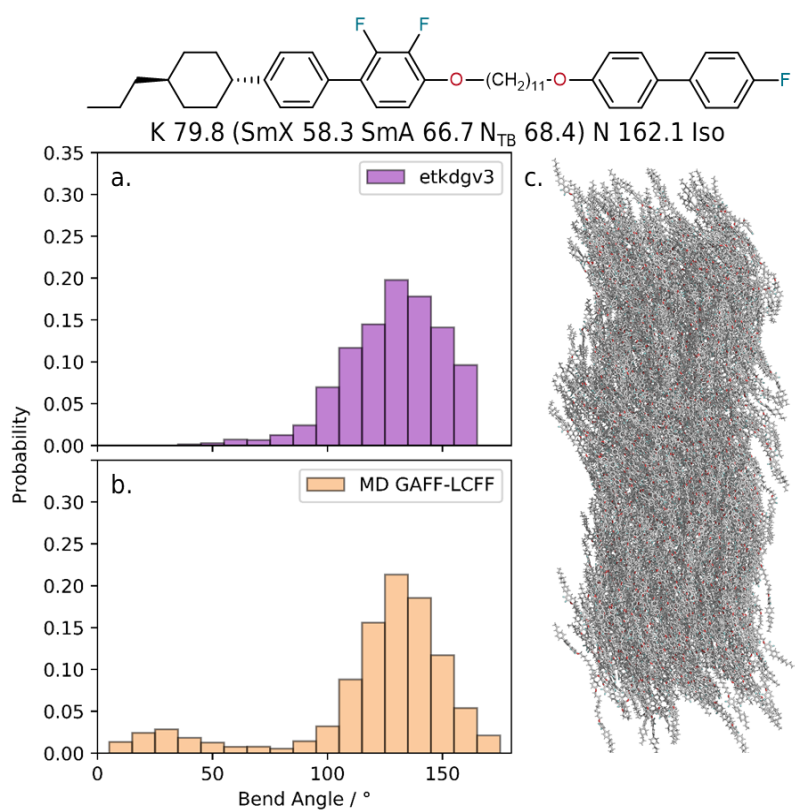

**Figure 8:** Molecular structure and transition temperatures (T, °C) of RM1081; [82] bend-angle distributions for RM1081 from (a) rules-based conformer generation using the ETKDGv3 method; (b) fully atomistic MD simulations at 325 K in the $N_{TB}$ phase; (c) instantaneous configuration of the $N_{TB}$ phase of RM1081 in an MD simulation at 325 K.

Thus far, we have demonstrated the feasibility of rapid assessment of molecular shape through the generation of conformer libraries using rules-based methods. Present results reiterate the need for a bent molecular shape in order for a given system to exhibit the $N_{TB}$ phase; in the broadest sense, the bend-angle should be in some optimal range, and that the system should not be entirely conformationally restricted but rather have some degree of rotational freedom. The analysis thus far has been *ex post facto*; we are only considering systems that have been synthesised, with reported experimental transition temperatures. It is also easy to see a case in which we apply this type of shape-analysis to evaluate synthetic targets, for example when changing/introducing linking groups.

We now move to consider the case where we use rules based methods to assess the shape of molecules designed algorithmically as a means to assess possible new $N_{TB}$ materials. Using the fragment-based generative molecular design algorithm described in ref [84], beginning with CB11CB as an input structure we generated ~ 10k candidate structures through fragment-based mutation. We filtered this list of candidate structures to remove molecules with fewer than 3 rings, molecules with rings smaller than 4 atoms, molecules without at least one 4-cyanobiphenyl, molecules with charges; this yielded circa 1k candidate molecules. We next calculated the bend-angle distribution for all molecules using the ETKDGv3 method and 128 conformers; from this, only molecules with a bend angle in the range 110-140 ° were retained. This list was then inspected visually, and some 'interesting' cases were selected and their geometry re-evaluated using the same ETKDGv3 method but with 2048 conformers.

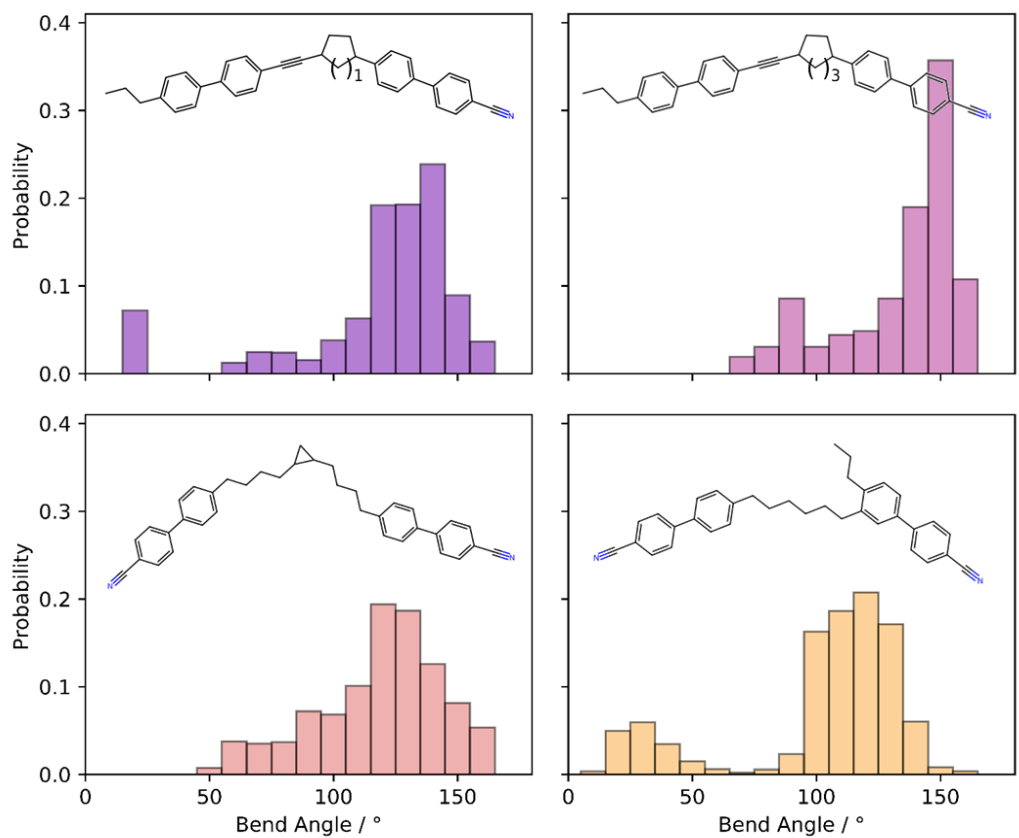

**Figure 9:** Bend-angle distributions of some algorithmically generated $N_{TB}$ like materials obtained *via* fragment based mutation of CB11CB.

Bent core-like materials with a central 1,3-cyclopentane or 1,4-cycloheptane are sufficiently non-linear that they possess average bend angles of ~ 130 °, and comparable distributions of angles to existing dimeric twist-bend nematic materials. The alicyclic ring ensures a degree of conformational freedom that is not found in conventional rigid aromatic core-units employed in bent-core LCs. We note that some bent-core LCs are known to exhibit the $N_{TB}$ phase already; the piperazine containing material UD68 is the best known example, and using the standard methodology described here we find it to have an average bend angle of 121 ° (**Figure SI-6**). The material with a central 1,2-disubstituted cyclopropane in the spacer is a structural isomer of CB11CB; the cyclopropyl ring enforces a bent-shape despite the even parity of the central spacer (with an average bend angle of ~ 130 °). In the last case, the hydrocarbon spacer has even parity and is bonded to the 3' position which gives a bent shape with an average bend angle of ~ 115 °, and despite having a distribution of bend angles which is somewhat broader than that of typical $N_{TB}$ materials. The manner of attachment of spacer to core is reminiscent of the λ-shaped trimers, [85,86] or χ-shaped tetramers [87] reported previously. From a computational perspective at least these structures appear to be reasonable synthetic targets, and indeed their synthesis would be anticipated to be relatively simple, although the capacity for mixed *para/meta* linked dimers to exhibit liquid crystalline phases is uncertain.

## Discussion

We have described the use of rules-based methods for rapidly assessing the molecular shape of semi-flexible liquid crystals. A lightweight GUI has been developed to enable these methods to be accessed easily, and we hope this will be of use to those working to develop/synthesise such materials.

We find that the fast conformer generation methods are generally adept at generating bend-angle distributions that compare favourably with MD simulations. Rules-based methods do not take into account the local environment, and the conformational preferences of flexible materials are probably related to the degree of orientational and/or positional order in a bulk phase. In the systems studied here, we find undersampling of hairpin conformations and oversampling of extended configurations, using fully atomistic MD simulations as a reference. However we feel the described methods are a significant improvement on the current the *de facto* standard of using a single conformer for molecular shape/geometry analysis, which neglects conformations through choice.

Upon re-evaluation of earlier data, we find that for a selection of CB7CB analogues the $N_{TB}$ transition temperature, scaled against the clearing point, correlates somewhat with bend angle for a given family of materials. Of course, molecules are not simple three-dimensional shapes; to ignore the propensity and capacity of different chemical functionality to generate specific mesophases is unwise. The software tools described herein are also used to explore the conformational landscapes of materials where systematic conformer searches would be prohibitively expensive.

Lastly, we show how conformational analysis can be integrated into a pipeline of algorithmic molecular development; using CB11CB as an input structure we use a fragment-based genetic algorithm to obtain new cyanobiphenyl-containing materials with bend-angles suitable for exhibiting this phase.

## Software availability

The GUI tool described herein is available from GitHub
https://github.com/RichardMandle/conformer

## Acknowledgements

RJM gratefully acknowledges funding from UKRI via a Future Leaders Fellowship, grant no. MR/W006391/1, ongoing support from Merck, and funding from the University of Leeds via a University Academic Fellowship. MD simulations and electronic structure calculations were undertaken on ARC4, part of the High Performance Computing facilities at the University of Leeds, UK.